\documentclass[prd,twocolumn,superscriptaddress,showpacs,preprintnumbers,nofootinbib,amsmath,amssymb,floatfix,aps]{revtex4-1}
\usepackage{graphicx}
\usepackage{subfigure}%
\usepackage{color}
\usepackage{isotope}

\graphicspath{{figs/}}

\begin{document}

\title{Inelastic Neutrino-Nucleus Interactions within the Spectral Function Formalism}
\author{Erica Vagnoni}
\email{vagnoni@fis.uniroma3.it}
\affiliation{INFN and Dipartimento di Matematica e Fisica, Universit\`a Roma Tre,
I-00146 Roma, Italy}
\author{Omar Benhar}
\email{omar.benhar@roma1.infn.it}
\affiliation{INFN and Dipartimento di Fisica, ``Sapienza'' Universit\`a di Roma, I-00185 Roma, Italy}
\affiliation{Center for Neutrino Physics, Virginia Tech, Blacksburg, Virginia 24061, USA}
\author{Davide Meloni}
\email{meloni@fis.uniroma3.it}
\affiliation{INFN and Dipartimento di Matematica e Fisica, Universit\`a Roma Tre,
I-00146 Roma, Italy}

\date{\today}%

\begin{abstract}
We report the results of a study of neutrino-carbon interactions at beam energies ranging between few hundreds 
MeV and few tens of GeV, carried out within the framework of the impulse approximation using a  realistic spectral function. 
The contributions of quasi elastic scattering, resonance production and deep inelastic scattering\textemdash consistently obtained, for 
first time, from a model based on a realistic description of the nuclear ground state\textemdash  are compared and analyzed.
\end{abstract}

%


\maketitle

Over the past decade, the broad effort aimed at improving the oversimplified description of neutrino-nucleus interactions
based on the Relativistic Fermi Gas Model (RFGM), has led to the development of a number of more advanced approaches, capable of
providing a fairly accurate description of part of the available data  \cite{Benhar05,Benhar:2006nr,GIBBU,coletti,PhysRevLett.98.242501,
Martini:2010ex,martini,Amaro:2010sd,PhysRevC.83.045501,nieves,Gran:2013kda,natalie,megias}. Most existing studies are restricted to the 
charged-current quasi-elastic (CCQE) sector, which makes the dominant contribution to the neutrino-carbon cross sections 
measured by the MiniBooNE Collaboration using a neutrino flux of mean energy $\langle E_\nu \rangle~\sim~800 \ {\rm MeV}$~\cite{miniboone_ccqe_2,AguilarArevalo:2010zc}. 
However, the interpretation of the signals relevant to ongoing and future experiments at higher neutrino energies, such as MINER$\nu$A \cite{minerva}, 
NO$\nu$A \cite{Ayres:2007tu} and DUNE \cite{DUNE} requires accurate predictions of the nuclear cross sections in inelastic channels. For example, 
at $E_\nu  \lesssim 2 \ {\rm GeV}$, corresponding to the peak energy of the  NO$\nu$A oscillated $\nu_e$ events, the total cross section is expected to 
receive comparable contributions from CCQE, resonance production and deep inelastic scattering (DIS) processes \cite{Ayres:2007tu}.

Theoretical calculations of the neutrino-nucleus cross section involve three main elements\textemdash the target initial and final states and the nuclear weak current\textemdash
whose {\em consistent} description in the broad kinematical region corresponding to neutrino energies between few hundreds MeV and few GeV  
poses severe difficulties. The initial state can be safely modeled within the non relativistic approximation,
independent of kinematics, whereas at large momentum transfer ${\bf q} = {\bf k} - {\bf k}^\prime$, the same approximation cannot
be used to describe either the nuclear final state, comprising at least one particle carrying momentum $\sim {\bf q}$, or the nuclear current operator,  
which depends explicitly on momentum transfer.

The impulse approximation (IA)\textemdash a detailed derivation of which can be found in Refs.~\cite{Benhar05,Benhar:2006wy}\textemdash provides a 
conceptual framework ideally suited to circumvent the above problem. The main tenet underlying this scheme is that, at large momentum transfer, nuclear interactions reduce to the
incoherent sum of elementary processes involving individual nucleons. As a consequence, nuclear and weak interaction dynamics are decoupled, and\textemdash to the extent to which the 
corresponding neutrino-nucleon cross section can be measured using hydrogen and deuterium targets\textemdash the formalism based on the IA can be used to describe
neutrino-nucleus scattering in any channels.

In this Letter, we report the results of the first comprehensive study of the neutrino-carbon cross section\textemdash including 
CCQE interactions, resonance production and  DIS\textemdash carried out within the IA using a realistic spectral function. 

The differential cross section of the process
\begin{align}
\nu_\mu + {^{12}{\rm C}}  \to \mu^- + X \ , 
\end{align}
in which a neutrino of four-momentum $k=(E_\nu,\bf k)$ scatters off a carbon nucleus producing a muon of four-momentum
$k^\prime=(E_\mu,{\bf k}^\prime)$, with the nuclear final state being undetected, can be written in the form
\begin{align}
\label{sigma:A}
\frac{d^2\sigma}{d\Omega_\mu dE_\mu}=\frac{G_F^2\,V^2_{ud}}{16\,\pi^2}\,
\frac{|\bf k^\prime|}{|\bf k|}\,L_{\mu\nu}\, W_A^{\mu\nu} \ ,
\end{align}
where $\Omega_\mu$ is the solid angle specified by the direction of the vector ${\bf k}^\prime$, $G_F$ is the Fermi constant and $V_{ud}$ is the element of the Cabibbo-Kobayashi-Maskawa (CKM) matrix  coupling $u$ and $d$ quarks.

The tensor $L_{\mu\nu}$ is completely determined by lepton kinematics, whereas the nuclear response to weak interactions is described by the tensor
\begin{align}
\label{hadronictensor}
W_A^{\mu\nu}= \sum_X \,\langle 0 | {J_A^\mu}^\dagger | X \rangle \,
      \langle X | J_A^\nu | 0 \rangle \;\delta^{(4)}(p_0 + q - p_X),
\end{align}
where  $|0\rangle$ and $|X\rangle$ denote the target ground state and the hadronic final state, 
carrying four momenta $p_0$ and $p_X$, respectively, $J_A^\mu$ is the nuclear weak current and 
the sum is extended to all hadronic final states.

The formalism of IA is based on the factorization {\em ansatz}, which amounts to replacing~\cite{Benhar05,Benhar:2006wy}
\begin{align}
\label{factorization}
|X\rangle \longrightarrow | x, {\bf p} \rangle \otimes |R, {\bf p}_R\rangle \ ,
\end{align}
where $|x, {\bf p} \rangle$ is the hadronic state produced at the electromagnetic vertex with  momentum ${\bf p}$,
while $|R, {\bf p}_R\rangle$ describes the recoiling nucleus, carrying momentum  ${\bf p}_R$.

It follows that Eq.\eqref{sigma:A} reduces to the simple and transparent form 
\begin{align}
\label{sigma:IA}
\frac{d^2\sigma_{IA}}{d\Omega_\mu dE_\mu}=
\int d^3k\,dE \,P({\bf k},E)\,\frac{d^2\sigma_{\nu N}}{d\Omega_\mu dE_\mu} \ ,
\end{align}
where the elementary $\nu N$ cross section\textemdash written in terms of five structure functions $W_i$\textemdash describes the interaction between the incoming 
neutrino and a {\em moving bound} nucleon, while the nuclear spectral function  
$P({\bf k},E)$\textemdash trivially related to the imaginary part of the two-point Green's function~\cite{BFF1,BFF2}\textemdash
yields the probability of removing a nucleon of momentum ${\bf k}$ from the target ground state, leaving
the residual nucleus with excitation energy $E$. 

Equation~\eqref{sigma:IA} clearly illustrates the potential of the formalism
based on the factorization {\em ansatz} of Eq.~\eqref{factorization}. 
Because the spectral function is an {\em intrinsic} property of the target ground state, it 
can be obtained from non relativistic nuclear many-body theory, and employed to carry out calculations of the nuclear cross section in any channels, provided the corresponding
$\nu$-nucleon cross section {\em in vacuum}  is known. The elementary cross section can be treated using the relativistic
formalism without any problems, 
nuclear medium effects being taken into account 
through the replacement~\cite{Benhar:2006wy}
\begin{align}
\label{omegatilde}
\omega = E_\nu - E_\mu \to {\widetilde \omega} = \omega + M_A  - E_{\bf p} - E_R \ ,
\end{align}
where $M_A$ is the target mass, while $E_{\bf p}$ and $E_R$ denote the energies of the 
hadronic state produced at the neutrino interaction vertex and of the recoiling nucleus, 
respectively.  Equation~\eqref{omegatilde} allows to account for 
the fact that, even though the weak interaction involves an individual nucleon, a fraction of the energy transfer in the scattering process goes into excitation 
energy of the spectator system. 

In the CCQE channel, characterized by the absence of pions in the final state, the relevant elementary interaction process is 
\begin{align}
\nu_\mu + n \to \mu^- + p \ , 
\end{align}
and the nucleon structure functions\textemdash involving a $\delta$-function constraining the mass of the hadronic final state to be equal to the proton mass, $m_p$\textemdash can be written in terms of the nucleon 
vector and axial-vector form factors. The former have been accurately measured in electron-proton 
and electron-deuteron experiments~\cite{Kelly2004,BBBA}, while the latter is usually written in the dipole form 
\begin{align}
F_A(Q^2) = g_A \left( 1 + Q^2/M_A^2 \right)^{-2} \ , 
\end{align}
with $Q^2 = - (k - k^\prime)^2$. The axial-vector coupling constant, $g_A=~-1.2761^{+14}_{-17}$, is known from neutron $\beta$-decay~\cite{g_A}, while the value of the axial mass 
is determined from elastic neutrino- and antineutrino-nucleon scattering, charged pion
electro-production off nucleons and muon capture on the proton~\cite{bernard,bodek2}.

The results reported in this Letter have been obtained using the state-of-the-art parametrization of the vector form factors of Ref.~\cite{BBBA}, and the dipole parametrization of the 
axial-vector form factor with $M_A = 1.03 \ {\rm GeV}$.

Conceptually, the generalization to describe resonance production, driven by elementary processes such as
\begin{align}
\label{delta++}
\nu_\mu + p \to \mu^- + \Delta^{++}  \to \mu^- + p + \pi^+ \ ,
\end{align}
where $\Delta^{++}$ denotes the $P_{33}(1232)$ nucleon resonance, only requires minor changes \cite{Benhar:2006nr}.
In this case, the $\nu N$ cross section involves  the matrix elements of the weak current describing the nucleon-resonance transitions.
As a consequence, the structure functions\textemdash which can still be written in terms of phenomenological vector and axial-vector form factors\textemdash  depend 
on both $Q^2$ and $W^2$, the squared invariant mass of the state $| x, {\bf p} \rangle$. In addition, the energy conserving $\delta$-function 
is replaced by a Breit-Wigner factor, accounting for the finite width of the resonance. 

Besides the prominent $P_{33}(1232)$ state, providing the largest contribution to the cross section, we have taken into account the three isospin $1/2$ 
states\textemdash  $D_{13}(1520)$, $P_{11}(1440)$, and $S_{11}(1535)$\textemdash comprised in the so-called second resonance region. 
The numerical results have been obtained using the parametrization of the structure functions described  in Refs~\cite{res1,res2,res3}. Within this approach, the vector form factors
are constrained by electroproduction data, while the axial couplings are extracted from the measured resonance decay rates, exploiting the Partially Conserved 
Axial Current (PCAC) hypothesis.   

From the observational point of view, Deep Inelastic Scattering (DIS) is associated  with hadronic final states comprising more than one pion.

In principle, the three nucleon structure functions determining the $\nu N$ cross section in the DIS  regime\textemdash $W_1$, $W_2$ and $W_3$\textemdash 
may be obtained combining measured neutrino and antineutrino scattering cross sections. However, as the available structure functions have been
extracted from {\em nuclear} cross sections (see, e.g., Ref.~\cite{CDHS}),
their use in {\em ab initio}  theoretical studies, aimed at identifying nuclear effects, entails obvious conceptual difficulties.

An alternative approach, allowing to obtain the structure functions describing DIS on isolated nucleons, can be developed within the conceptual
framework of the quark-parton model, exploiting the large database of accurate DIS data collected using charged lepton beams and  hydrogen and deuteron targets (see, e.g., Ref.~\cite{roberts}).
Within this scheme,  the function $F_2^{\nu N} = \omega W_2$, where $W_2$ is the structure function of  an isoscalar nucleon,
can be simply related to the corresponding structure function extracted from electron scattering data, $F_2^{e N}$ through\footnote{For the sake of simplicity, here, and in what follows, we will ignore the contributions of $s$ and $c$ quarks.}
\begin{align}
F_2^{\nu N}(Q^2,x) = \frac{18}{5} \  F_2^{e N}(Q^2,x)  \ ,
 \label{DIS1}
\end{align}
where $x$ is the Bjorken scaling variable. In addition, the relation
\begin{align}
\label{DIS2}
x F_3^{\nu N}(Q^2,x)  & = x \ [ \ u_{\rm v}(Q^2,x)  + d_{\rm v}(Q^2,x)  \ ] \ ,
\end{align}
where $F_3^{\nu N} = \omega W_3$ and  
$u_{\rm v}$  and $d_{\rm v}$ denote the valence quark distributions, implies
\begin{align}
\label{DIS3}
x F_3^{\nu N}(Q^2,x)  & = F_2^{eN}(Q^2,x) \\
\notag
& - 2 x \  [ \overline{u}(Q^2,x) + \overline{d}(Q^2,x)]   \ .
\end{align}
Using Eqs.~\eqref{DIS1}-\eqref{DIS3} and the Callan-Gross relation~\cite{roberts},  linking  $F_1^{\nu N} =  m W_1$ to $F_2^{\nu N}$,  one can readily obtain all the relevant weak
structure functions from the existing parametrizations of the measured electromagnetic structure function and of the antiquark distributions $\overline{u}$ and $\overline{d}$ (see, e.g., Ref~.\cite{GRV98}).
Alternatively, the quark and antiquark distributions can be also used to obtain the structure function $F_2^{e N}$ from
\begin{align}
\notag
F_2^{e N}(Q^2,x)  = x \ \frac{5}{18}  [ \  & u(Q^2, x)  + \overline{u}(Q^2,x)  \\
 & + d(Q^2,x) + \overline{d}(Q^2,x) \ ]  \ .
\label{DIS0}
\end{align}
In this work, we have used Eqs.\eqref{DIS1}-\eqref{DIS0} and the parton distributions of Ref.~\cite{GRV98}, which are available for $Q^2~\geq~Q^2_{\rm min}~=~0.8$~GeV$^2$. At lower values of $Q^2$, we have assumed
the parton distributions to be the same as at $Q^2=Q^2_{\rm min}$. 

Note that the above procedure rests on the tenet, underlying the IA scheme, that the elementary neutrino-nucleon interaction is {\em not} affected by the
presence of the nuclear medium, the effects of which are accounted for with the substitution of Eq.\eqref{omegatilde}. While this assumption is strongly supported by electron-nucleus scattering data in the quasi elastic channel, showing no evidence of medium modifications of the nucleon vector form factors, it has to be mentioned that analyses of neutrino DIS data are often
carried out within a conceptually different approach, allowing for medium modifications of either the nucleon structure functions \cite{petti,haider}, or of the parton distributions entering their definitions \cite{kumano}.

\begin{figure}[h!]
\vspace*{-.15in}
\begin{center}
\includegraphics[scale=0.64]{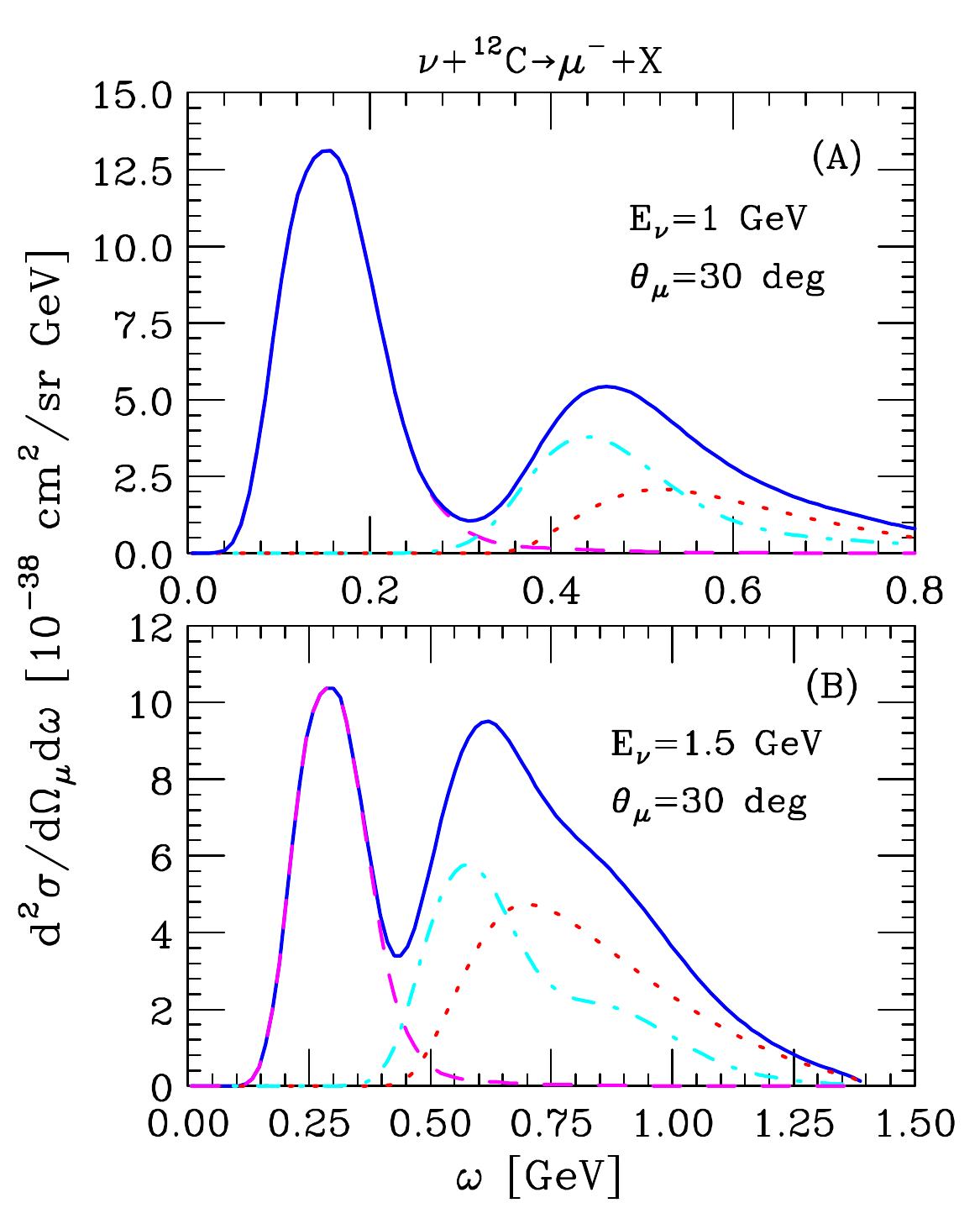}
\end{center}
\vspace*{-.30in}
\caption{Double-differential cross section of the  scattering process $\nu_\mu + {^{12}C} \to \mu^- + X$ at fixed muon emission angle $\theta_\mu = 30 \ {\rm deg}$, and beam energies 
$E_\nu = $ 1 GeV (A) and 1.5 GeV (B), displayed as a function of $\omega = E_\nu - E_\mu$. The dashed, dot-dash and dotted lines 
correspond to CCQE scattering, resonance production and DIS, respectively. The sum of the three contributions is represented 
by the full line.}
\label{d2sigma}
\end{figure}

The results of calculations of the electron-nucleus cross sections have provided ample evidence that the 
approach based on IA and the spectral function formalism, {\em involving no adjustable parameters}, is capable to deliver a quantitative description of the 
double-differential electron-nucleus cross sections\textemdash measured at fixed beam energy and electron scattering angle\textemdash
in both the qualsielastic and inelastic sectors~\cite{electron,LDA}. Figure~\ref{d2sigma} shows the results of the extension of these analyses
to the case of  neutrino-carbon interactions. The calculations have been carried out using the spectral function of Ref.~\cite{LDA} and
setting the muon emission angle to $\theta_\mu = 30 \ {\rm deg}$. Comparison between panels (A) and (B), corresponding to $E_\nu =$ 1 and 1.5 GeV, respectively, illustrates 
how the relative weight of the different reaction mechanisms changes with increasing neutrino energy.

\begin{figure}[h!]
\begin{center}
\includegraphics[scale=0.64]{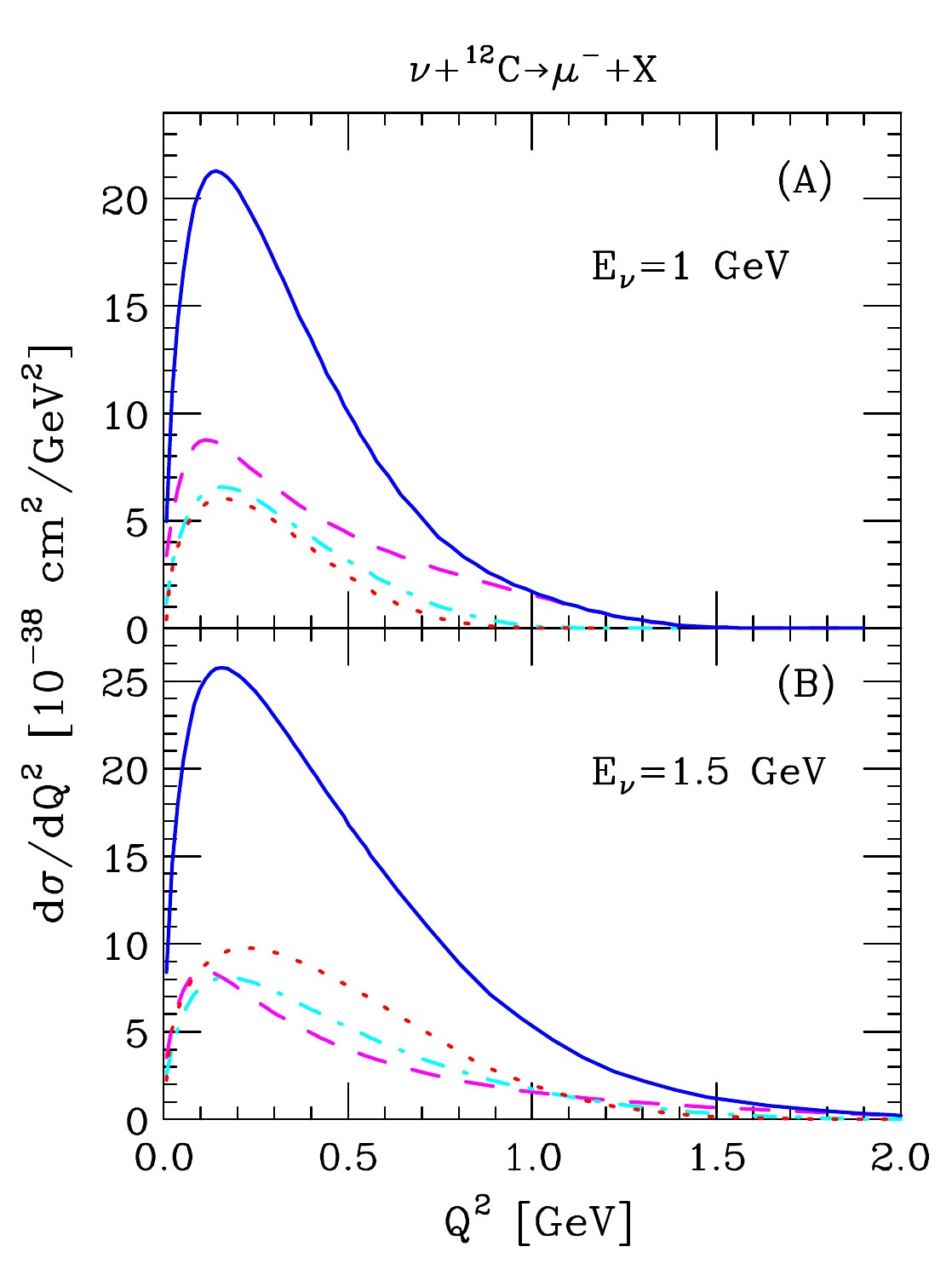}
\end{center}
\vspace*{-.35in}
\caption{$Q^2$-distribution of the process $\nu_\mu + {^{12}C} \to \mu^- + X$ at fixed neutrino energy $E_\nu$ = 1 Gev (A) and 1.5 GeV (B). The meaning of the lines is the 
same as in Fig.~\ref{d2sigma}.}
\label{Q2dist}
\end{figure}

The $Q^2$-distributions, obtained from the double-differential cross section of Fig.~\ref{Q2dist} by integrating over $\cos \theta_\mu$, are displayed in Fig.~\ref{Q2dist}.
At both $E_\nu =$ 1 and 1.5 GeV, the full $d\sigma/dQ^2$, corresponding to the solid line, exhibits a pronounced maximum at $Q^2 \lesssim 0.2 \ {\rm GeV}^2$. 

\begin{figure}[t!]
\begin{center}
\includegraphics[scale=0.65]{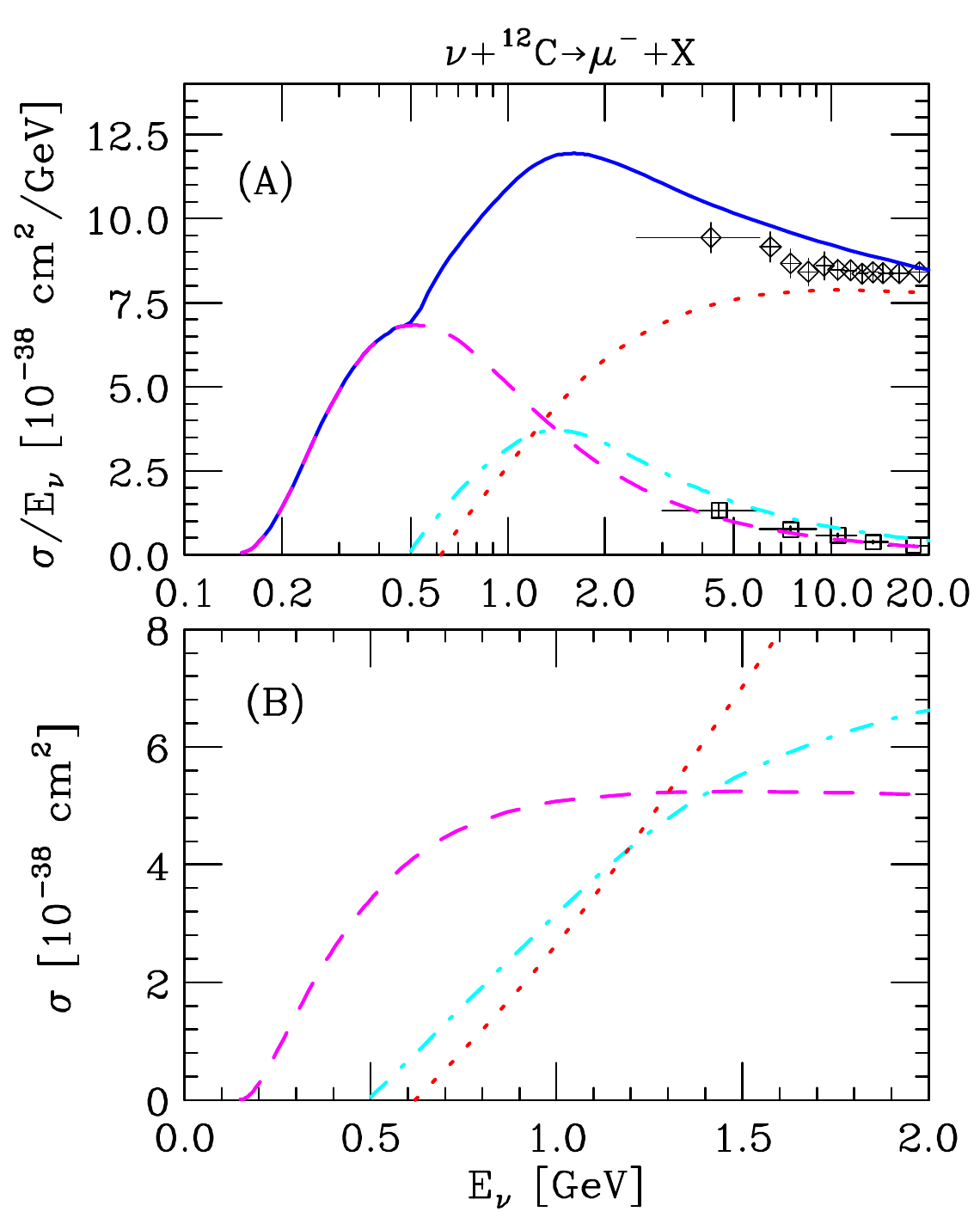}
\end{center}
\vspace*{-.30in}
\caption{Total cross section of the reaction $\nu_\mu + {^{12}C} \to \mu^- + X$ as a function of neutrino energy. The dashed, dot-dash and dotted lines of panel (A) 
represent the contributions of CCQE, resonance production and DIS processes. Panel (B) shows the $E_\nu$-dependence of the ratio $\sigma/E_\nu$.
The meaning of the dashed, dot-dash and dotted lines is the same as in panel (A). The full line corresponds to the sum of the three contributions.
Diamonds and squares represent the data of Refs.\cite{NOMAD_tot,NOMAD_CCQE}, respectively.}
\label{sigmatot}
\end{figure}

Finally, integration over $Q^2$ yields the total cross section, $\sigma$, whose behavior as a function of the neutrino energy $E_\nu$ is illustrated in Fig.~\ref{sigmatot}. 
Panels (B) and (A) show $\sigma$ and the ratio $\sigma/E_\nu$, respectively, as well as
the contributions corresponding to the CCQE, resonance production, and DIS channels. It is apparent that, while at $E_\nu \lesssim 0.8$ GeV CCQE interactions dominate, 
the inelastic cross section rapidly increases with energy. At  $E_\nu \approx 1.3$ GeV, the contributions arising from the three reaction channels turn out to be about the same. 

For comparison, in panel (B) we also report, as diamonds, the $\nu_\mu$-carbon total cross section measured by the NOMAD collaboration \cite{NOMAD_tot}.
It turns out that, while the energy-dependence of the data at $E_\nu \gtrsim 10$ GeV is well reproduced by our prediction of the DIS contribution, represented by the dotted line,  
the results of the full calculation, corresponding to the solid line, sizably exceed the measured cross section. In view of the fact that the CCQE cross section obtained from the NOMAD 
data of Ref.~\cite{NOMAD_CCQE}, shown by the open squares, turns out to be in close agreement with the results of our calculations, 
this discrepancy is likely to be ascribed to double counting between resonance production and DIS contributions, which are very hard to identify in a truly model independent fashion.  

In conclusion, we have carried out a calculation based on the IA and the spectral function formalism, in which the contributions of 
CCQE processes, resonance production and DIS are taken into account, for the first time, in a fully consistent fashion.  
The present implementation of the factorization scheme does not take into account the occurrence 
of processes involving more than one nucleon\textemdash such as those in which the neutrino couples to nuclear Meson-Exchange-Currents (MEC)\textemdash  
as well as  final state interactions (FSI) between the nucleon participating in the weak interaction process and the spectator particles. 
The inclusion of MEC contributions to the neutrino-nucleus cross section is believed to be needed to explain the flux-integrated double-differential cross 
section measured by the MiniBooNE collaboration~\cite{martini,nieves,megias}, while the understanding of FSI is required, e.g., to determine the nuclear transparency to the 
hadrons produced at the interactions vertex~\cite{transparency}.

Theoretical studies of the electron-carbon cross section provide convincing evidence that MEC contributions can be consistently included in the spectral function formalism, through a 
generalization of the factorization ansatz \cite{MEC1,MEC2}, while FSI corrections in the quasi elastic channel are understood at quantitative level \cite{FSI1,FSI2}. 
Note, however, that FSI {\em do not} affect the CCQE total cross section shown in Fig.~\ref{sigmatot}.

The emerging picture suggests that the approach based on spectral functions strongly constrained by {\em both inclusive and exclusive} electron-nucleus scattering data, such as those derived 
in Ref.~\cite{LDA}, has the potential to describe both elastic and inelastic neutrino-nucleus interactions at the level of accuracy required to face the outstanding challenges of neutrino physics.


\begin{acknowledgments}
The authors are grateful to Artur M.~Ankowski and Camillo~Mariani for countless illuminating 
discussions. 
The work of E.V. and D.M. was supported by INFN through grant WSIP.
The work of O.B. was supported by INFN through grant MANYBODY.

\end{acknowledgments}

\end{document}